\documentclass[preprint,a4paper,showpacs,amssymb]{revtex4}
\usepackage{graphicx}
\begin{document}


\title{Polarization effects in the channel of an organic field-effect transistor}
\author{H. Houili, J.D. Picon}
\affiliation{\it{Laboratoire d'Opto\'{e}lectronique des Materiaux Mol\'{e}culaires, STI-IMX-LOMM, station 3, Ecole Polytechnique
F\'{e}d\'{e}rale de Lausanne, CH-1015, Lausanne, Switzerland}.}

\author{M.N. Bussac}
\affiliation{\it{Centre de Physique Th\'{e}orique, UMR-7644 du Centre National de la Recherche Scientifique, Ecole Polytechnique,
F-91128 Palaiseau Cedex, France}.}

\author{L. Zuppiroli}\email[Corresponding author: ]{libero.zuppiroli@epfl.ch}
\affiliation{\it{Laboratoire d'Opto\'{e}lectronique des Materiaux Mol\'{e}culaires, STI-IMX-LOMM, station 3, Ecole Polytechnique
F\'{e}d\'{e}rale de Lausanne, CH-1015, Lausanne, Switzerland}.}

\begin{abstract}
We present the results of our calculation of the effects of dynamical coupling of a charge-carrier to the electronic polarization and the field-induced lattice displacements at the gate-interface of an organic field-effect transistor (OFET). We find that these interactions reduce the effective bandwidth of the charge-carrier in the quasi-two dimensional channel of a pentacene transistor by a factor of two from its bulk value when the gate is a high-permittivity dielectric such as $\left(\textrm{Ta}_{2}\textrm{O}_{5}\right)$ while this reduction essentially vanishes using a polymer gate-insulator. These results demonstrate that carrier mass renormalization triggers the dielectric effects on the mobility reported recently in OFETs.
\end{abstract}
\pacs{72.80.Le, 73.40.Qv, 32.10.Dk, 77.22.Ch}
\keywords{}

\maketitle

\section{INTRODUCTION}
Organic field-effect transistors (OFETs) with high charge-carrier mobilities
are essential components for high-speed organic electronic applications. For
this reason, it is of crucial importance to unravel the origin of charge
transport in these devices. Several experimental studies have shown that the
difference in the charge-carrier transport observed in bulk organic
semiconductors and within the quasi-two dimensional channel near the
gate-insulator interface in OFETs is associated with the effects of disorder
and interfacial traps \cite{PMB04,DML04,KSB05,DMN05}. More intriguing is a
growing body of evidence demonstrating the strong influence of the
dielectric permittivity of the gate-insulator on the charge-carrier mobility
in OFETs \cite{VOL03,SBI04}. Veres et al. \cite{VOL03} have studied the case
of triarylamine polymer transistors in which measured mobilities well below $10^{-2}\text{ cm}^{2}/\text{V s }$can be unambiguously attributed to hopping
between localized states. On the other hand, Stassen et al. \cite{SBI04}
obtained much higher values, from $1$ to $20\text{ cm}^{2}/\text{V s}$, in
single-crystalline rubrene transistors in which conduction is more
intrinsic. A surprising finding in both cases is the drastic decrease of the
mobilities as the dielectric constant of the gate-insulator is increased
systematically using different dielectric materials. Such observation in
devices governed by vastly different charge-transport mechanisms strongly
suggests an effect due to the interactions of the charge carriers with the
gate-dielectric. These interactions can lead to the renormalization of the
bare mass of the charge carriers in the conducting channel of OFETs,
providing an explanation for the observed dielectric effects on the mobility
as reported in Refs. \cite{VOL03,SBI04}.

In this work, we present a theoretical calculation of the renormalized band
mass of charge carriers in the conducting channel of OFETs by looking at two
important interactions experienced by charges at the interface of an organic
semiconductor with a dielectric material. The first one, purely electronic
in origin, is the image force due to the polarization discontinuity at the
gate-dielectric interface. The second involves the Coulomb interaction of
the charge carriers with surface polar phonons of the dielectric. To
illustrate the basic concepts of our calculations and their general
applicability to any organic semiconductor, we choose pentacene as a model
system since its lattice structure, bandwidth, polarizabilities, and
Huang-Rys factors are well-known compared with other organic semiconductors.
We find that, in pentacene, the cumulative effect of the electronic and
surface-lattice polaronic interactions is to reduce the effective bandwidth
by a factor of two as the relative static dielectric constant of the gate
materials is varied from $1$ to $25$. Within the tight-binding model, such a
decrease in the transfer integral can be viewed as a concomitant increase in
the effective mass by the same factor. We will suggest that such
renormalization of the band properties triggers the effects reported in
Refs. \cite{VOL03,SBI04}.

\section{TIME SCALES}

The non-interacting band properties of a perfect pentacene crystal along all
crystallographic directions have been calculated by Cheng et al. \cite{CSS03}. For the bare transfer integral, $J$, between molecules along the direction
of easy propagation in the $\left( a,b\right) $-plane, they obtain $J=100\text{ meV from which one gets }h/J\simeq 4\times 10^{-14}$ s$\text{ }$as the characteristic time for in-plane Bloch-wave formation. The corresponding transfer time in the perpendicular $c$-axis direction, $h/J_{\bot }$, is thirty times longer than in the plane. Thus, in the presence of scattering which substantially reduces the Bloch-wave lifetime, the carrier motion is essentially two-dimensional.

The above considerations allow for the classification of the various interactions experienced by charge carriers in organic semiconductors. For fast interactions with characteristic times shorter than $h/J$, the charge can be assumed to be located on a single molecular site. In pentacene, this is the situation encountered during the interaction of the carrier with the electronic polarizability of the medium or in intramolecular charge-transfer as well as the coupling with intramolecular carbon stretching vibrations with frequencies around $1360\text{ cm}^{-1}$. Since fast interactions arise prior to the formation of the Bloch-wave, they have the effect of dressing the charge with a polarization cloud or a lattice deformation cloud. Slow interactions, on the other hand, have characteristic times much longer than $h/J$. They act directly on the Bloch-wave or the localized state. Such is the case for interaction of the charge carrier with low-energy
intermolecular thermal phonons and librations which, in many cases, can be
considered as static with respect to the two-dimensional band motion. These
interactions scatter the Bloch-wave or localize the electronic states when
the disorder they introduce is large enough. The interaction of the charge
with the surface polar phonons of the gate insulator occurs in the intermediate time scale regime. An interesting discussion of time scales can also be found in the first chapter of the book by Silinsh and \v{C}\'{a}pek \cite{Silinsh94}.

Because they dress the charge with a polarization cloud or lattice
deformation, fast processes lead to a renormalization of the bare transfer
integrals $J$ and $J_{\bot }$ and consequently increase the effective mass
along all crystal directions. The case involving electron-phonon
interactions has been discussed by several authors, including Appel \cite{Appel} and Davydov \cite{Davydov}. The purely electronic effects were treated by three of us in an earlier work \cite{BPZ04} in which we calculated the renormalization effect due to the electronic polarizability in the bulk of the organic semiconductor. In this work, these calculations will be extended to the situation encountered by carriers at the gate-dielectric interface in OFETs. We shall treat both electronic and lattice effects. The Fr\"ohlich surface polaron at the oxide surface was already studied by Kirova and Bussac \cite{KB03} for an isotropic organic crystal. The entire problem will be revisited here for the two-dimensional
layer of the anisotropic crystal. The slow processes involving low-energy
phonons, librations and other quasi-static or static sources of scattering
and localization will be discussed and treated elsewhere.

We have to emphasize that, because they are faster than $10^{-14}\text{ s}$,
the polarization processes studied here involve only the high-frequency
dielectric response of the materials constituting the interface and not the
usual low frequency (static) permittivity.

For the sake of clarity and to highlight the important aspects of this work,
we summarize the major results of our calculations in the next section. The
details of these calculations will be given in separate appendices.

\section{RESULTS}

In rubrene- or acene-based organic field-effect transistors, the interface
with the oxide or polymer gate-insulator involves the highly conducting $\left(a,b\right)$-plane of the organic semiconductor. Our calculation of
the variation of the effective transfer integral, $J_{IV}$, for conduction
in this plane is shown in Fig.\ref{fig:1} as a function of the static
dielectric constant of the gate-insulator. In the results of Fig.\ref{fig:1}, the charge-carrier is assumed to be located on the first monolayer close
to the gate-interface. In the case where the charge is on the second
monolayer, we find that there is basically no effect of the dielectric on
the transfer integral which then takes the bulk value. These results are
obtained from the combined effects of four different interactions which have
been treated separately according to their time scales from the fastest to
the slowest as discussed below.

\begin{enumerate}
\item \textit{Electronic polarization }$\left( J^{I}\right)$. The dynamical
renormalization of the carrier motion due to electronic polarization is
different in the bulk and at the interface. At an interface, an image field
is generated which is attractive when the high-frequency dielectric
constant, $\epsilon _{\infty }$, of the gate-insulator is greater than that
of the organic semiconductor as is the case for oxides, and is repulsive
otherwise as in the case of polymers or air-gap insulators. This image field
is typically of the same order of magnitude as the applied gate fields to
which it is added or subtracted. The magnitude of the image potential under
usual experimental conditions is displayed in Fig.\ref{fig:2} as a function
of the distance $z$ from the interface. At large distances, the classical
expression for the image field holds and the image potential is written as, 
\begin{equation}
E_{p}\left( z\right) -E_{p}\left( \infty \right) =-\frac{e^{2}}{16\pi
\epsilon _{0}z}\left( \frac{\epsilon _{\infty ,2}-\epsilon _{\infty ,1}}{%
\epsilon _{\infty ,1}\left( \epsilon _{\infty ,2}+\epsilon _{\infty
,1}\right) }\right)
\end{equation}
where $\epsilon _{\infty ,1}$ and $\epsilon _{\infty ,2}$ are the
high-frequency dielectric constants of the semiconductor and the
gate-insulator, respectively. However, corrections to this expression
associated with lattice effects show up close to the interface.

The increase in the carrier effective mass due to the electronic
polarization cloud is slightly different in the bulk or when it crosses the
interface. The details of these calculations are given in Appendix \ref{appen:A}. They lead to a renormalized intermolecular charge-transfer integral $J^{I}<J$.

\item \textit{Electronic displacement }$\left( J^{II}\right)$. The strong
dependence of the mobility and the effective mass with the dielectric
permittivity of the gate-insulator seen in experiments, as well as large
corrections to the bulk electronic polarization energy near the interface
shown in Fig.\ref{fig:2}, suggest that the first two monolayers next to the
dielectric interface dominate the charge transport particularly in the
presence of a significant gate field \cite{DML04}. At even higher fields,
one may expect that not only is the charge localized on the first monolayer
but its electronic wavefunction is squeezed towards the part of the molecule
closer to the insulator. This displacement of the charge distribution on the
molecule is also a fast process, controlled by the transfer integral $t_{/\!/}\sim 1\text{ eV}$ within the molecule which is more than an order of
magnitude larger than $J^{I}$. This fast process decreases further the
transfer integral to $J^{II}$. The recent semi-empirical quantum-chemistry
calculation performed by Sancho-Garc\'{\i}a et al. \cite{SHB03} suggests
that this effect is completely negligible in pentacene where the
charge-carrier distribution remains perfectly centered even at very high
fields ($100\text{ MV/cm}$). In this case $J_{II}=J_{I}$. The final result
of our calculation presented in Fig. \ref{fig:1} was established in the
frame of this hypothesis that we consider as the most reliable at the
moment. Nevertheless, in Appendix \ref{appen:B} another approach is
presented which shows that much larger effects would be expected if the
charge-carrier distribution on the molecule is allowed to be displaced by
values of a few angstroems at high gate fields of $10\text{ MV/cm}$.

\item \textit{Intramolecular vibrations} $\left( J^{III}\right) $.
Intramolecular vibrations close to $1360\text{ cm}^{-1}$ in pentacene are
strongly coupled to the carrier because they change the $\pi $-alternation
typical of conjugated molecules. Because these atomic motions are faster
than the electronic polaron motion defined by the renormalization transfer
integral $J^{II}$, they also contribute to a further reduction of $J^{II}$
by a constant factor of $0.75$, independent of the distance from the
interface as well as the applied field $\left( J^{III}\simeq
0.75J^{II}\right) $. Appendix \ref{appen:C} reviews how this factor is
calculated.

\item \textit{Fr\"ohlich polaron at the oxide surface }$\left( J^{IV}\right)$. Oxides are polar materials. Thus, the infrared-active phonon modes which
modulate the metal-oxide bonds are strongly coupled to charge carriers
sitting at their surface. In aluminum oxide, for instance, the most active
mode of this kind is situated at $46\text{ meV }$\cite{STH00}. This value is
of the same order of magnitude as the effective, in-plane, transfer integral
renormalized upon corrections due to the above-mentioned interactions.

The construction of a lattice deformation cloud in the oxide is the object
of Appendix \ref{appen:D}. The polarization interaction energy of the charge
with this cloud causes further attraction of the charge to the surface and
to the subsequent increase of the effective mass. The calculation is
performed in the intermediate coupling regime because here the coupling
parameter which controls the process, $\alpha _{\text{eff}}\left( z\right)$, defined in Appendix \ref{appen:D}, is of the order of unity in the first
monolayer, and of the order of $0.1$ in the second one.
\end{enumerate}

The total binding energy of the carrier in the presence of a gate-insulator
includes both the electronic and surface polaron effects arising from the
electronic image force potential and the lattice deformation potential at
the interface associated with the above interactions. Within a tight-binding
model, these effects are incorporated into a renormalized effective transfer
integral $J_{IV}$ given by, 
\begin{equation}
J_\textrm{IV}=J\left( \frac{J^{I}}{J}\right) \left( \frac{J^{II}}{J^{I}}%
\right) \left( \frac{J^{III}}{J^{II}}\right) \left( \frac{J^{IV}}{J^{III}}%
\right)
\end{equation}
Table \ref{tab:3} provides a summary of all these factors in the order of
increasing time-scale characterizing each interaction.

\section{CONCLUSION}
The experimental results have clearly shown that the mobilities obtained in
organic field-effect transistors are much larger in devices built with an
air-gap or a polymer gate-insulator \cite{PMB04,VOL03,SBI04} than those
using high-permittivity oxide gate dielectrics. Our theoretical results
discussed above provide some insight into the origin of these effects.

Given that the bandwidth is four times the transfer integral as suggested in
the quantum chemistry calculations of Ref. \cite{SKB05} and starting from a
value of $390\text{ meV }$\cite{CSS03}, the effective bandwidth becomes $231\text{ meV in bulk pentacene }$\cite{BPZ04}. It is reduced to $\text{155 meV}$ in an OFET with an aluminum oxide gate insulator $\left( \text{Al}_{2}\text{O}_{3}\right) $, to $\text{146 meV}$ close to a $\text{Ta}_{2}\text{O}_{5}$ interface and $\text{144 meV}$ close to a $\text{Ti}\text{O}_{2}$ interface, while the bulk value of $231\text{ meV}$ is recovered close to a parylene gate insulator.

For a given disorder potential in the organic semiconductor, localization
effects roughly scale with the reciprocal bandwidth. Consequently, important
bandwidth reductions enhance all localization effects and consequently
decrease the mobility. It is important to note that in the present work the
existence of a lattice polaron in the bulk of pentacene was considered as
unlikely. Intramolecular vibrations are too fast to produce a polaron in a
perfect pentacene crystal and their effect has been studied in Appendix \ref{appen:C}. Intermolecular phonons and librations with energies $\hbar \omega$ of the order of $10\text{ meV}$ are not coupled enough to the carrier to
produce a well-defined lattice polaron, with a renormalized transfer
integral of the order of $50\text{ meV}$ originating from the initial
calculation of Cheng et al. \cite{CSS03}. However, when the transfer
integral experiences a further reduction close to the gate interface as
shown in Fig. \ref{fig:1}, then the intrinsic lattice polaron can be
excited, as suggested in a recent work \cite{HB04}. A calculation is in
progress to clarify this point. Moreover, an interface is rarely perfect
from a structural point of view, and the fact that, due to significant
internal field present at the interface, the carrier is constrained to probe
more closely all the interface disorder, enhancing the localization effects.
Even "perfect" interfaces can be "intrinsic" sources of localization. In
general, the electric dipole lattice induced by a charge carrier in the
organic semiconductor is incommensurate with the dipole lattice induced in
the gate material. In a perfect structure, the incommensurability of the
electronic potential can open gaps in the semiconductor density of states.
When the gate dielectric is not a single-crystal but a disordered structure,
the same effect creates an electronic disorder which can be one of the
sources of localization.

The effective bandwidth enters the mobility law but does not define it
entirely. In disordered polymer channels, Veres et al. \cite{VOL03} have
established a link between the effective width of the Gaussian distribution
of electronic states and the mobility of the carrier which jumps from state
to state according to the Gaussian Disorder Model \cite{Bassler93}. In more
ordered systems such as single-crystalline rubrene OFETs, a theoretical work
is in progress to establish a quantitative link between the transfer
integral and the mobility. It will elucidate the role of thermal, low-energy
phonons which are, on the one side, sources of localization and on the other
the key to the adiabatic diffusion of the carriers in the channel.

\begin{acknowledgments}
The authors acknowledge the financial support of the Swiss Federal Science Foundation under contract number 200020-105156.
\end{acknowledgments}

\appendix
\section{\label{appen:A}Electronic polaron at the interface}
The electronic polarization of a molecular crystal, $i.e.$ the formation of
induced dipoles $\mathbf{d}$ on the neutral molecules of the crystal in the
field of an extra-charge, occurs on a time scale of $\tau \sim
10^{-15}-10^{-16}$ s $\left( h/\tau \sim 2\text{ eV}\right) $. This is much
faster than the carrier lifetime on a molecule which is typically $\tau
_{J}\sim h/J\sim 4\times 10^{-14}$ s with $J\simeq 100\text{ meV,}$ allowing
one to treat the carrier as stationary on a molecular site during the
polarization process.

\subsection{Bulk}

With the above approximation, we have calculated the polarization energy of
a point-charge in bulk pentacene crystal to be $E_{p}\simeq -1.5\text{ eV}$  
\cite{BPZ04} using the method of self-consistent polarization field to the
dipolar approximation \cite{Silinsh94}. The pentacene crystal structure and
parameters are given in Tab.\ref{tab:param}. The molecules involved in the
discrete calculation are assumed to be inside a spherical cluster of radius $%
R$. The polarization energy varies linearly with the inverse radius of the
cluster. The polarizabilities of the molecules along the $\left(
L,M,N\right) $-directions were adjusted so that the polarization energy at
large $R$ matches the continuum limit $\left( 1-\epsilon _{\text{eff}%
}^{-1}\right) e^{2}/4\pi \epsilon _{0}R$, where $\epsilon _{\text{eff}}$ is
the effective dielectric constant in an anisotropic structure \cite{BM79}.
The convergence is fast and the calculated polarization energies for
anthracene, $E_{p}\simeq -1.26\text{ eV}$, and pentacene, $E_{p}\simeq -1.5%
\text{ eV}$, are in good agreement with recently reported values \cite{TS03}.

As the charge carrier moves from one molecule to another, the polarization
energy changes. In fact, when the electronic polarization energy and the
induced-dipole distribution vary from site $n$ to site $n+h$ due to
different local crystal environment, the carrier will encounter a resistance
to its motion. This leads to a renormalization of the bare transfer integral 
$J$. We introduced the parameter $S_{0\text{ }}$\cite{BPZ04} as a way of
describing quantitatively the polarizability effect on the transfer
integral, $i.e.$ $J^{I}=J\exp \left( -S_{0}\right) $. The quantity $S_{0}$
is closely related to the relative amplitudes and directions of the induced
dipoles and is given through the relations \cite{BPZ04}, 
\begin{eqnarray}
S_{0}\left( h\right)  &=&\frac{1}{2}\sum_{i}\sum_{l=1}^{3}\left(
X_{i,l}\left( n\right) -X_{i,l}\left( n+h\right) \right) ^{2}  \label{eqa:A1}
\\
X_{i,l} &=&\frac{d_{i,l}}{2\sqrt{\varepsilon \alpha _{ll}}}
\end{eqnarray}%
where $\alpha _{ll}$ are the components of the polarizability, and $%
\varepsilon $, of the order of one electron-volt, is the energy difference
between the ground state and the first-excited state of the neutral molecule 
\cite{BPZ04}. The sites considered here, $n$ and $n+h$, are the next-nearest
neighbor along the crystal direction $\mathbf{d}_{2}=-\frac{1}{2}\mathbf{a}+%
\frac{1}{2}\mathbf{b}$.

\subsection{Interface}

We now extend the above calculation to the case of charge carrier near the
interface of pentacene with the dielectric insulator. In this case, the
polarization cloud extends partly into the semiconductor and partly into the
dielectric. For simplicity, we modelled the dielectric insulator as a cubic
lattice. The Clausius-Mosotti relation is then applied to obtain the
electronic polarizability of the molecules of the dielectric assuming a
given high frequency permittivity, 
\begin{eqnarray}
\frac{\epsilon _{\infty ,2}-1}{\epsilon _{\infty ,2}+2} &=&\frac{1}{%
3\epsilon _{0}}N\alpha _{e}  \label{eq:C-M} \\
N &=&\frac{\rho }{M}N_{A}
\end{eqnarray}%
Here $M$ and $\rho $ are the molar mass and the density of the dielectric
and $N_{A}$ is the Avogadro number. The summation in Eq.\ref{eqa:A1} is then
extended to include the induced dipoles in the semiconductor and the
dielectric. For reasons of symmetry, the molecules involved in the discrete
calculations are taken inside a cylindrical cluster on both sides of the
interface. Beyond the cluster approximation, a continuum contribution is
added to the polarization energy using standard electrostatics \cite%
{Jackson99}. The polarization energy as a function of the distance from the
interface is shown in Fig.\ref{fig:2} for both the discrete and the
continuous limits.

\section{\label{appen:B}Squeezing of the charge distribution at the interface}

The self-trapping interfacial polarization electric field (see Appendix \ref%
{appen:A}) and the applied gate-field localize the excess charge on the
pentacene molecule by pulling and squeezing the wavefunction towards or away
from the interface depending on the relative magnitudes of the dielectric
constants of pentacene and the gate-insulator. To get the average position
and the spatial extent of the wavefunction of the excess charge, we used a
model in which the pentacene molecule is considered as a pair of conjugated
chains of eleven sites, each separated by a distance $l$ that we treat in
the tight-binding approximation. The basis wavefunction on each site, $\psi
_{n},$ with $n=1,..,11$ satisfies the following set of equations, 
\begin{equation}
E_{0}\psi _{n}=-t_{/\!/}\left( \psi _{n+1}+\psi _{n-1}\right) -nlqF\psi
_{n}-V\left( nl\right) \psi _{n}
\end{equation}%
with the boundary conditions $\psi _{0}=\psi _{12}=0$ imposed at the two
ends of the pentacene molecule. Here $V$ is the image force potential, $F$
is the applied gate-field, and $t_{/\!/}$ is the intramolecular transfer
integral. Upon solving this system of equations, we obtain the groundstate
wavefunction as, 
\begin{equation}
|\Psi _{0}\rangle =\sum_{i=1}^{11}\psi _{n}|n\rangle ,
\end{equation}%
The spatial dependence of the average position of the excess charge on the
chain will be given by, 
\begin{equation}
\langle n\rangle =\sum_{n=1}^{11}n\left\vert \psi _{n}\right\vert ^{2}
\label{eq:mean}
\end{equation}

Since the pentacene molecule is nearly perpendicular to the interface, the presence of interfacial fields localizes the charge at the extremity of the molecule. This has the effect of changing the thickness of the channel and, as a consequence, the bandwidth for charge propagation. Indeed, the transfer integral in acene crystals increases with the number of aromatic rings from naphtalene to pentacene \cite{CSS03,BCS02}. This can be related to the spatial extent of the excess charge on the acene molecule. The charge extent $\sigma _{n}$ is defined as the width containing $80\%$ of the charge density. We have chosen as in Appendix \ref{appen:A} the direction of higher transfer integral which determines the bandwidth. Therefore, the variation of the charge extension under the effect of the image force and the gate-field, will induce a corresponding change in the transfer integral of the pentacene molecules as shown in Fig. \ref{fig:3}. For the calculation of the excess charge extension in the above model, the intramolecular transfer integral $t_{/\!/}$ was set to $1$eV. The values of the gate and image force fields are those found in Appendix \ref{appen:A}. However, a semi-empirical quantum chemistry calculation published recently \cite{SHB03} has shown that, probably due to exchange interactions between the carrier and the $\pi$-electrons in the pentacene molecule, the extra-charge distribution is not pulled and squeezed in the direction of the interface field by the extent predicted by the above model. Thus in the final result of Fig. \ref{fig:1} this effect has not been included. Figure \ref{fig:3} shows what would the final result be in the case where the squeezing calculated above in this appendix would indeed be effective.

\section{\label{appen:C}Band narrowing due to intramolecular vibrations}
We now refer to the strong coupling of the extra charge to the
intramolecular vibrations at $1360\text{ cm}^{-1}$ in pentacene. The
coupling constant has been determined in acenes both experimentally and
theoretically in Ref. \cite{CMS02}. The calculation of the band narrowing due
to phonons has been presented by several authors \cite{Appel,Davydov}. Here,
we present a simplified version for dispersionless intramolecular vibrations.

The electron-phonon Hamiltonian of interest is written as, 
\begin{equation}
H=\sum_{n,h}-J^{II}a_{n+h}^{+}a_{n}+\sum_{n}\hbar \omega _{0}b_{n}^{+}b_{n}-%
\sqrt{E_{B}/\hbar \omega _{0}}\left( b_{n}^{+}+b_{n}\right) a_{n}^{+}a_{n}
\end{equation}
where $n$ represents the molecular sites; $a_{n}^{+}$, $a_{n}$, $b_{n}^{+}$,
and $b_{n}$ are the electron and phonon operators respectively. $g=\sqrt{%
E_{B}/\hbar \omega _{0}}$ is the usual coupling parameter. Since the phonon
frequency $\hbar \omega _{0}$ is larger than the transfer integral $J^{II}$,
we look for a variational solution of the Hamiltonian. The trial
wavefunction is of the form, 
\begin{equation}
|\psi _{n}\rangle =\sum_{n}u_{n}|n\rangle \otimes |\chi _{n}\rangle 
\end{equation}
where 
\begin{equation}
|\chi _{n}\rangle =\exp \left( X_{n}^{\ast }b_{n}-X_{n}b_{n}^{+}\right)
|0\rangle 
\end{equation}
describes the intramolecular vibrations of the molecule $n$ on which the charge
is located. The variationnal parameter $X_{n}$ is determined by minimizing
the energy 
\begin{equation}
E=\langle \psi |H|\psi \rangle 
\end{equation}
Then 
\begin{eqnarray}
E &=&-\sum_{n,h}J^{II}\exp \left( \frac{-|X_{n}|^{2}+|X_{n+h}|^{2}}{2}%
\right) u_{n+h}^{+}u_{n}  \nonumber \\
&&-\sqrt{E_{B}/\hbar \omega _{0}}\left( X_{n}+X_{n}^{\ast }\right)
|u_{n}|^{2}+\sum_{n}\hbar \omega _{0}|X_{n}|^{2}
\end{eqnarray}
We find $X_{n}=\sqrt{E_{B}/\hbar \omega _{0}}$ and the corresponding
transfer integral, 
\begin{equation}
J^{III}=J^{II}\exp \left( -E_{B}/\hbar \omega _{0}\right) 
\end{equation}
When thermal phonons are taken into account the thermal average value of the
energy yields a temperature correction factor in the transfer integral given
by,
\begin{equation}
J^{III}=J^{II}\exp \left( -\frac{E_{B}}{\hbar \omega _{0}}\coth \left( \frac{%
\hbar \omega _{0}}{2k_{B}T}\right) \right) 
\end{equation}
In the present case, the temperature effect is completely negligible and the
electron-phonon binding energy is just $-E_{B}$ per molecular site.

Taking the values in bulk pentacene from Ref. \cite{CMS02} with $E_{B}=45%
\text{ meV}$, we obtain $J^{III}=0.75J^{II}$. Close to the interface the
band narrowing due to intramolecular vibrations is rather insensitive to the
interfacial field. This is related to the fact that the phonon frequency and
coupling constants are approximately the same in the acene series \cite{CMS02}.

\section{\label{appen:D}The Fr\"ohlich surface polaron}

When a charge carrier is generated by the field effect at the interface
between a molecular semiconductor and a gate insulator, it interacts with
the surface phonons of the dielectric. This effect has been studied in Ref. \cite{KB03} for an isotropic 3D molecular crystal in the adiabatic limit. Here,
we consider this interaction in the case of a pentacene crystal where the
carrier motion is essentially two-dimensional and for moderate
electron-phonon coupling. Indeed, the residence time in a monolayer $\tau
\sim \hbar /J_{\perp }$, with $J_{\perp }\sim 5\text{ meV}$ is much larger
than the time to polarize the dielectric given by $2\pi /\omega _{s}$, where 
$\hbar \omega _{s}=46\text{ meV}$ in $\text{Al}_{2}\text{O}_{3}$ \cite{STH00}%
. However, the phonon frequency $\hbar \omega _{s}$ is of the same order of
magnitude as the effective in-plane transfer integral $J^{III}\sim 60\text{
meV}$. The frequency $\omega_s$ is given by the following formula:
\begin{equation}
\omega_{s}^{2}=\frac{1}{2}\left(\omega_{L}^{2}+\omega_{T}^{2}\right),
\end{equation}
where $\omega_{L}$ and $\omega_{T}$ are the frequencies of the bulk longitudinal and transverse phonons. \cite{Sak72} The values of $\omega_s$ for the different dielectrics are reported in Tab. \ref{tab:2}

The electron-phonon interaction involving a charge in a particular monolayer
of the crystal at a distance $z>0$ from the interface is given by \cite%
{Silinsh94}, 
\begin{equation}
H_{\text{e-ph}}=\sum_{q}\frac{e}{\sqrt{q}}\sqrt{\frac{\pi \hbar \omega _{s}}{%
S\epsilon ^{\ast }}}\sum_{n}e^{-qz}e^{i\mathbf{q}\cdot \mathbf{n}a}\left(
b_{q}+b_{-q}^{+}\right) |\psi _{n}|^{2}
\end{equation}%
where $|\psi _{n}|^{2}$ is the charge density at site $\mathbf{n}a=\left(
n_{x}a,n_{y}a\right) $. Here $b_{q}$ and $b_{-q}^{+}$ are the annihilation
and creation operators of the surface phonons in the gate material, $\omega
_{s}$ their frequency, and $S$ the surface area of the interface.

If $\epsilon _{1}\left( \omega \right) $ is the dielectric susceptibility of
the molecular crystal, $\epsilon _{\infty ,2}$ and $\epsilon _{0,2}$ the
high and low frequency limits of the dielectric permittivity, the coupling
constant $1/\epsilon ^{\ast }$ is given by, 
\begin{equation}
\frac{1}{\epsilon ^{\ast }}=\frac{\epsilon _{1}-\epsilon _{\infty ,2}}{%
\epsilon _{1}\left( \epsilon _{1}+\epsilon _{\infty ,2}\right) }-\frac{%
\epsilon _{1}-\epsilon _{0,2}}{\epsilon _{1}\left( \epsilon _{1}+\epsilon
_{0,2}\right) }
\end{equation}%
and the total Hamiltonian becomes, 
\begin{equation}
H=-J^{III}a^{2}\frac{p^{2}}{\hbar ^{2}}+\sum_{q}\hbar \omega
_{s}b_{q}^{+}b_{q}+H_{\text{e-ph}}
\end{equation}%
where $\mathbf{p}$ is the momentum of the charge carrier. Following Ref. \cite{LLP53}, we introduce the total momentum of the system which is a constant
of motion of the total Hamiltonian,
\begin{equation}
\mathbf{P}=\sum_{q}\hbar \mathbf{q}b_{q}^{+}b_{q}+\mathbf{p}
\end{equation}
We can transform the total Hamiltonian $H$ to $H^{\prime }$ through the
unitary transformation $\hat{S}$, so that $H^{\prime }$ no longer contains
the charge coordinates, 
\begin{equation}
H^{\prime }=\hat{S}^{-1}H\hat{S}
\end{equation}%
with, 
\begin{equation}
\hat{S}=\exp \left[ i\left( \mathbf{P}-\sum_{q}b_{q}^{+}b_{q}\mathbf{q}%
\right) \cdot \mathbf{n}a\right] 
\end{equation}%
We obtain thus, 
\begin{eqnarray}
H^{\prime } &=&\sum_{q}\hbar \omega _{s}b_{q}^{+}b_{q}+\sum_{q}V_{q}\left(
z\right) \left( b_{q}+b_{-q}^{+}\right) \left[ \mathbf{P}/\hbar
-\sum_{q}b_{q}^{+}b_{q}q\right] J^{III}a  \nonumber \\
&&+\left[ \mathbf{P}/\hbar -\sum_{q}b_{q}^{+}b_{q}\right] J^{III}a^{2}
\end{eqnarray}%
where 
\begin{equation}
V_{q}\left( z\right) =\frac{e}{\sqrt{q}}\sqrt{\frac{\pi \hbar \omega _{s}}{%
S\epsilon ^{\ast }}}e^{-qz}
\end{equation}%
With the phonon frequency $\hbar \omega _{s}$ comparable to the effective
transfer integral $J^{III}$, the adiabatic approximation is not applicable.
However, the dimensionless parameter $\alpha _{\text{eff}}$ which describes
the strength of the electron-phonon coupling decreases with the distance $z$
to the interface as 
\begin{equation}
\alpha _{\text{eff}}=\frac{e^{2}}{8\pi \epsilon _{0}\epsilon ^{\ast }a}\frac{%
\exp \left( -2\pi z/a\right) }{\sqrt{\hbar \omega _{s}J^{III}}}\equiv \alpha
e^{-2\pi z/a}
\end{equation}%
In our case, $\alpha _{\text{eff}}$ is of the order of $1$ for a charge
carrier located in the first monolayer. Then, we use a variational method to
describe the interaction of the dressed charge carrier with the dielectric
phonons \cite{LLP53}. Introducing a second unitary transformation, 
\begin{equation}
\hat{U}=\exp \left( \sum_{q}b_{q}^{+}f_{q}-b_{q}f_{q}^{\ast }\right) 
\end{equation}%
where $f_{q}$ will be chosen to minimize the energy 
\begin{eqnarray}
E &=&\frac{\mathbf{P}^{2}}{\hbar }a^{2}J^{III}+\sum_{q}\left(
V_{q}f_{q}+V_{q}^{\ast }f_{q}^{\ast }\right) +J^{III}\left(
\sum_{q}|f_{q}|^{2}q^{2}a^{2}\right) ^{2}  \nonumber \\
&&+\sum_{q}|f_{q}|^{2}\left[ \hbar \omega _{s}+J^{III}\left( q^{2}a^{2}-2%
\mathbf{q}\cdot \frac{\mathbf{P}}{\hbar ^{2}}a^{2}\right) \right] 
\end{eqnarray}%
We then find, 
\begin{equation}
f_{q}=-\frac{V_{q}^{\ast }}{\hbar \omega _{s}+J^{III}}\left[ q^{2}a^{2}-2%
\frac{\mathbf{q}\cdot \mathbf{P}}{\hbar }a^{2}\left( 1-\eta \right) \right] 
\end{equation}%
where $\eta $ satisfies the implicit equation 
\begin{equation}
\eta{\bf P}=\frac{\sum_q \vert V_q\vert^2 \hbar {\bf q}}{\hbar\omega_s+J^{III}\left[q^2a^2-2\frac{{\bf q}\cdot{\bf P}} {\hbar}a^2\left(1-\eta\right)\right]}
\end{equation}
The carrier binding energy is obtained as $E_{b}=-\alpha I_{1}\left(
z\right) \hbar \omega _{s}$ and the effective mass is $m^{\ast
}/m=J^{III}/J^{IV}=1+2\alpha I_{2}\left( z\right) $. As long as $%
J^{III}P^{2}a^{2}/\hbar ^{2}$ is small ($\lesssim \hbar \omega _{s}$), we
may obtain $E\left( P^{2}\right) $ to first order in an expansion in powers
of $\left( \frac{J^{III}}{\hbar \omega _{s}}\frac{P^{2}a^{2}}{\hbar ^{2}}%
\right) $. On doing so, one readily gets 
\begin{equation}
E=-\alpha I_{1}\left( z\right) \hbar \omega _{s}+\frac{P^{2}a^{2}}{\hbar ^{2}%
}\frac{J^{III}}{\left[ 1+2\alpha I_{2}\left( z\right) \right] }
\end{equation}%
where 
\begin{equation}
I_{1}\left( z\right) =\int_{0}^{\pi \sqrt{J^{III}/\hbar \omega _{s}}}\frac{dy
}{1+y^{2}}\exp \left( -\frac{2z}{a}\sqrt{\frac{\hbar \omega _{s}}{J^{III}}}
y\right)g\left(y\right) 
\end{equation}%
and 
\begin{equation}
I_{2}\left( z\right) =\int_{0}^{\pi \sqrt{J^{III}/\hbar \omega _{s}}}\frac{%
y^{2}dy}{\left( 1+y^{2}\right) ^{3}}\exp \left( -\frac{2z}{a}\sqrt{\frac{%
\hbar \omega _{s}}{J^{III}}}y\right)g\left(y\right)
\end{equation}%
Here the term given by
\begin{equation}
g\left(y\right)=\frac{\sinh\left[{y\sqrt{\frac{\hbar\omega_s}{J^{III}}}%
\frac{\ell}{a}}\right]}{y\sqrt{\frac{\hbar\omega_s}{J^{III}}}\frac{\ell}{a}}
\end{equation}
accounts for the finite extension of the charge distribution on the molecule of length $\ell$.

\newpage

\section*{Table and figure captions}

\begin{list}{}{}
\item{\bf Figure 1}: The largest transfer integral for charge propagation in pentacene close to the interface with the gate insulator as a function of the relative static permittivity of the dielectric. The charge carrier is located on the first monolayer close to the interface. The high frequency dielectric constant $\epsilon_{\infty,2}$ for the gate insulators is: $2.65$ for parylene C; $2.37$ for $\textrm{Si}\textrm{O}_2$; $3.1$ for $\textrm{Al}_{2}\textrm{O}_3$; $4.12$ for $\textrm{Ta}_{2}\textrm{O}_5$; and $6.5$ for $\textrm{Ti}\textrm{O}_2$.
\item{\bf Figure 2}: The image potential $E_{p}\left(z\right)$ at the interface is given in the continuous approximation and in our lattice model. At large distances from the interface the electronic polarization energy in the bulk is recovered.
\item{\bf Figure 3}: These are the same results as in Fig. \ref{fig:1} but calculated with another set of data as explained in Appendix \ref{appen:B}. The largest transfer integral for charge propagation in pentacene close to the interface with the gate insulator is plotted as a function of the relative static permittivity of the dielectric. The charge-carrier is assumed to be located on the first monolayer. The inset depicts the case where the charge-carrier is on the second monolayer. These curves cumulate all the effects calculated in this work. In contrast to the results of Fig. \ref{fig:1} we have assumed here that the carrier distribution is pulled away or squeezed towards the interface according to the calculation in Appendix \ref{appen:B}. This effect is enhanced with respect to the semi-empirical model of Ref. \cite{SHB03}. Here the effective mass depends even on the gate field. At present, the corresponding values calculated in Fig. \ref{fig:1} are considered more reliable.
\item{\bf Table 1}: Crystal constants of pentacene [R. B. Campbell et al., Acta Cryst. {\bf 14}, 705 (1961)].
\item{\bf Table 2}: Values of the surface phonon frequencies used in the calculations of Appendix \ref{appen:D} together with the coupling constant $\alpha$ (see text).
\item{\bf Table 3}: Reduction factors of the transfer integral through the steps discussed in the text. Cases of bulk pentacene and interfaces with vacuum and different dielectric oxides are shown along with the time scales characterizing each process. Here the second effect which is related to charge displacement on the molecule was excluded (see text).
\end{list}

\newpage

\begingroup
\squeezetable
\begin{table}[!h]
\begin{tabular}{@{\hspace{1cm}}l@{\hspace{1cm}}l@{\hspace{1cm}}l@{\hspace{1cm}}l@{\hspace{1cm}}}\hline \hline
Crystal constants (\AA) & $a$ & $b$ & $c$ \\ \hline
&7.9 & 6.06 & 16.01 \\ \hline
Crystal constants (deg) & $\alpha$ & $\beta$ & $\gamma$ \\ \hline
& 101.9 & 112.6 & 85.8 \\ \hline\hline
Dielectric constant\footnote{\scriptsize{E.V. Tsiper and Z.G. Soos, Phys. Rev. B {\bf 68}, 085301 (2003).}} & $L$ & $M$ & $N$ \\ \hline
& 5.336 & 3.211 & 2.413 \\ \hline \hline
\end{tabular}
\caption{}
\label{tab:param}
\end{table}
\endgroup

\begingroup
\squeezetable
\begin{table}[!h]
\begin{tabular}{l|c|c}\hline \hline
 & $\omega_{s}\left(\textrm{cm}^{-1}\right)$ & $\alpha$ \\ \hline
$\textrm{Si}\textrm{O}_2$\footnote{\scriptsize{J. Huml\'i{\v c}ek, A. R\"oseler, Thin Solid Films {\bf 234}, 332 (1993)}} & 480 & 2.84 \\ \hline
$\textrm{Al}_{2}\textrm{O}_3$\footnote{\scriptsize{M. Schubert, et al., Phys. Rev. B {\bf 61}, 8187 (2000)}} & 386 & 5.24 \\ \hline
$\textrm{Ta}_{2}\textrm{O}_5$\footnote{\scriptsize{E. Franke, et al., J. Appl. Phys. {\bf 88}, 5166 (2000)}} & 390 & 6.17 \\ \hline
$\textrm{Ti}\textrm{O}_2$\footnote{\scriptsize{R. Sikora, J. Phys. Chem. Solids {\bf 66}, 1069 (2005)}} & 280 & 4.84 \\ \hline
\hline
\end{tabular}
\caption{}
\label{tab:2}
\end{table}
\endgroup

\begingroup
\squeezetable
\begin{table}[!h]
\begin{tabular}{l|c|c|c|c}\hline \hline
 & Molecular & Intramolecular & Surface phonons & final result \\ 
 & polarization & charge vibration & & \\ \hline
Typical time scales & $\sim 10^{-15}$ s & $\sim 2\times10^{-14}$ s & $\sim 8\times10^{-14}$ s & \\ \hline \hline
Reduction factors & $J^{I}/J$ & $J^{III}/J^{I}$ & $J^{IV}/J^{III}$ & $J^{IV}/J$ \\ \hline
Pentacene bulk & $0.79$ & $0.75$ & $1$ & $0.593$ \\ \hline
Pentacene/vacuum & $0.76$ & $0.75$ & $1$ & $0.57$ \\ \hline
Pentacene/$\textrm{SiO}_2$ & $0.77$ & $0.75$ & $0.815$ & $0.47$ \\ \hline
Pentacene/$\textrm{Al}_{2}\textrm{O}_3$ & $0.77$ & $0.75$ & $0.69$ & $0.398$ \\ \hline
Pentacene/$\textrm{Ta}_{2}\textrm{O}_5$ & $0.77$ & $0.75$ & $0.65$ & $0.375$ \\ \hline
Pentacene/$\textrm{TiO}_2$ & $0.77$ & $0.75$ & $0.64$ & $0.37$ \\ \hline
\hline
\end{tabular}
\caption{}
\label{tab:3}
\end{table}
\endgroup

\begin{figure}[!h]
\begin{center}
\includegraphics[]{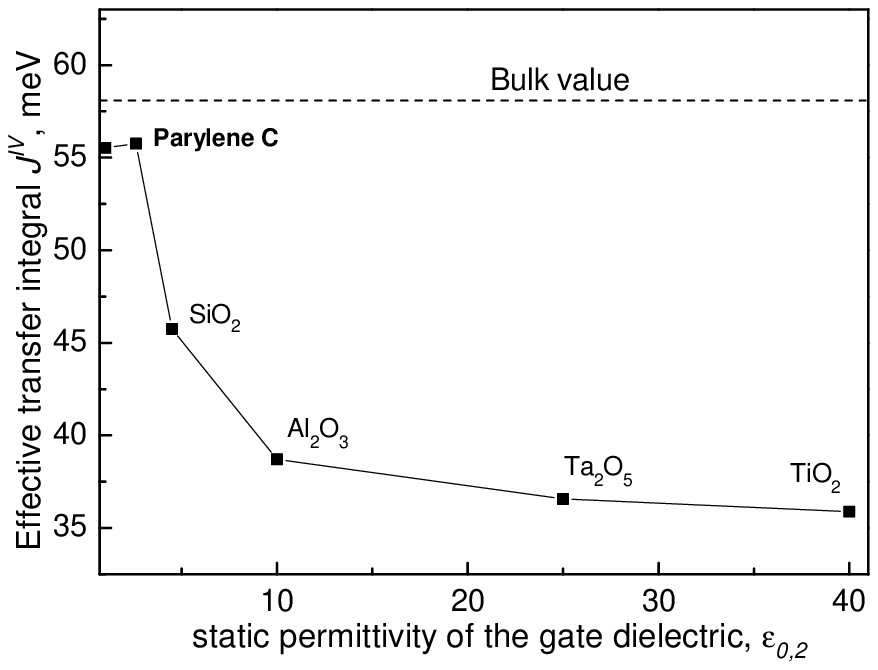}
\caption{}
\label{fig:1}
\end{center}
\end{figure}

\begin{figure}[!h]
\begin{center}
\includegraphics[]{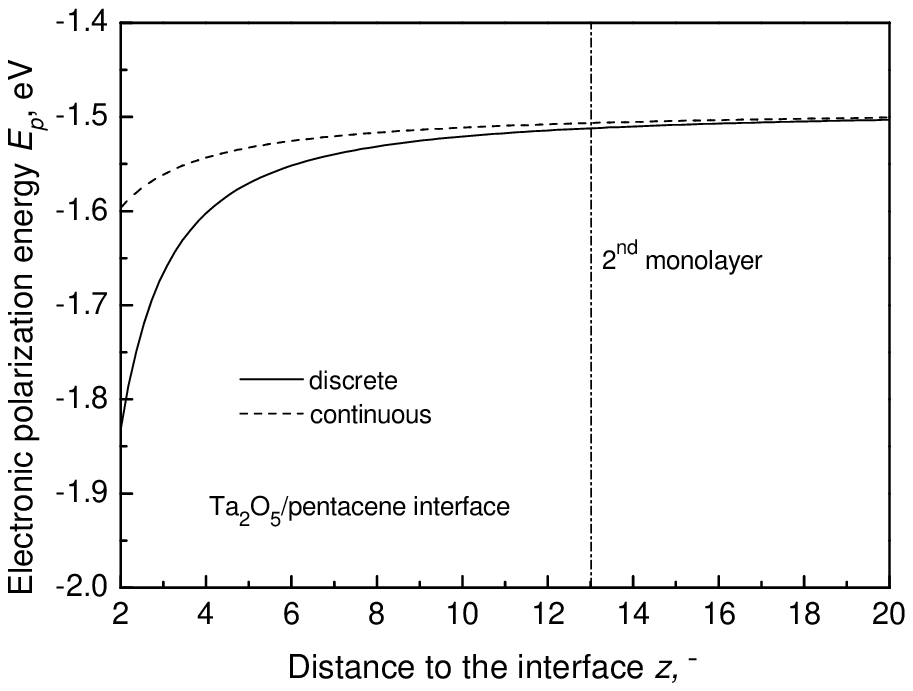}
\caption{}
\label{fig:2}
\end{center}
\end{figure}

\begin{figure}[!h]
\begin{center}
\includegraphics[]{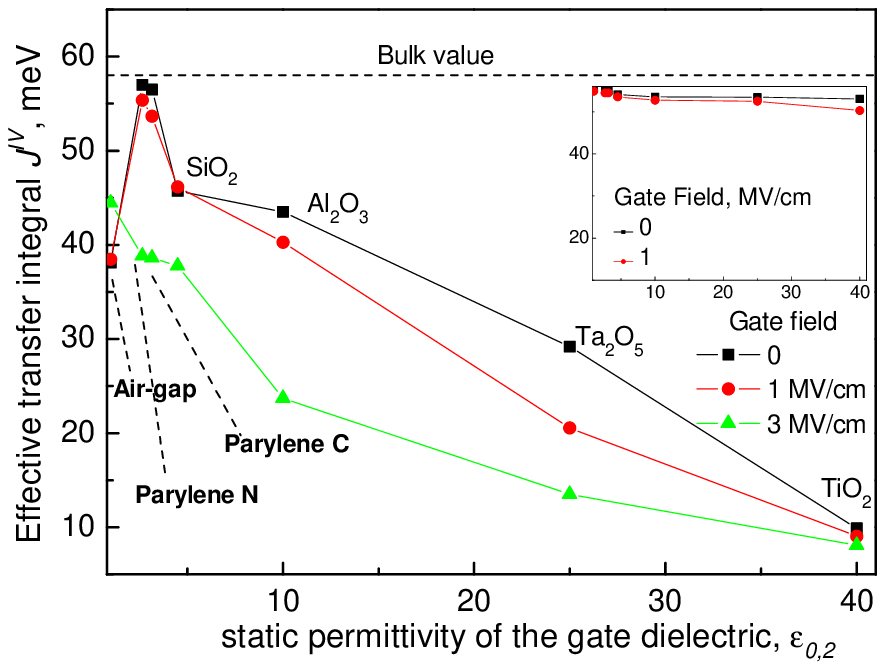}
\caption{}
\label{fig:3}
\end{center}
\end{figure}


\begin{thebibliography}{999}
\bibitem{PMB04} V. Podzorov, E. Menard, A. Borissov, V. Kiryukhin, J. A. Rogers, and M. E. Gershenson, Phy. Rev. Lett. {\bf 93}, 086602 (2004).
\bibitem{DML04} F. Dinelli, M. Murgia, P. Levy, M. Cavalleni, F. Biscarini, and D. M. de Leeuw, Phys. Rev. Lett. {\bf 92}, 116802 (2004).
\bibitem{KSB05} J. H. Kang, D. A. da Silva Filho, J. L. Bredas, and X.-Y. Zhu, Appl. Phys. Lett. {\bf 86}, 152115 (2005).
\bibitem{DMN05} M. Daraktchiev, A. von M\"uhlenen, F. N\"uesch, M. Schaer, M. Brinkmann, M. N. Bussac, and L. Zuppiroli, New Journal of Physics {\bf 7}, 133 (2005).
\bibitem{VOL03} J. Veres, S. D. Ogier, S. W. Leeming, D. C. Cupertino, and S. M. Khaffaf, Adv. Func. Mater. {\bf 13}, 199 (2003).
\bibitem{SBI04} A. F. Stassen, R. W. I. de Boer, N. N. Iosad, and A. F. Morpurgo, Appl. Phys. Lett. {\bf 85}, 3899 (2004).
\bibitem{CSS03} Y. C. Cheng, R. J. Silbey, D. A. da Silva Filho, J. P. Calbert, J. Cornil, J. L. Bredas, J. Chem. Phys. {\bf 118}, 3764 (2003).
\bibitem{Silinsh94} E. A. Silinsh and V. \v C\'apek, Organic Molecular Crystals (AIP Press, New York, 1994).
\bibitem{Appel} J. Appel, Solid Stat. Phys. {\bf 25}, 193 (1968).
\bibitem{Davydov} A. S. Davydov, Th\'eorie du Solide (Edition Mir, 1980).
\bibitem{BPZ04} M. N. Bussac, J. D. Picon, and L. Zuppiroli, Europhys. Lett. {\bf 66}, 392 (2004).
\bibitem{KB03} N. Kirova, and M. N. Bussac, Phys. Rev. B {\bf 68}, 235312 (2003).
\bibitem{SHB03} J. C. Sancho-Garc\'ia, G. Horowitz, J. L. Br\'edas, and J. Cornil, J. Chem. Phys. {\bf 119}, 12563 (2003).
\bibitem{STH00} M. Schubert, T. E. Tiwald, and C. M. Herzinger, Phys. Rev. B {\bf 61}, 8187 (2000).
\bibitem{SKB05} D. A. da Silva Filho, E. G. Kim, and J. L. Br\'edas, Adv. Mater. {\bf 17}, 1072 (2005).
\bibitem{HB04} K. Hannewald, and P. A. Bobbert, Phys. Rev. B {\bf 69}, 075212 (2004); K. Hannewald, V. M. Stojanovi\'c, J. M. Schellekens, P. A. Bobbert, G. Kresse, and J. Hafner, {\it ibid.} {\bf 69}, 075211 (2004).
\bibitem{Bassler93} H. B\"assler, Phys. Stat. Sol. B {\bf 175}, 11 (1993).
\bibitem{BM79} P. J. Bounds, and R. W. Munn, Chem. Phys. {\bf 44}, 103 (1979).
\bibitem{TS03} E. V. Tsiper, and Z. G. Soos, Phys. Rev. B {\bf 68}, 085301 (2003).
\bibitem{Jackson99} J. D. Jackson, Classical Electrodynamics (Wiley \& Sons, New York, 1999), p. 154.
\bibitem{BCS02} J. L. Bredas, J. P. Calbert, D. A. da Silva Filho, and J. Cornil, Proc. Natl. Acad. Sci. {\bf 99}, 5804 (2002).
\bibitem{CMS02} V. Coropceanu, M. Malagoli, D. A. da Silva Filho, N. E. Gruhn, T. G. Bill, and J. L. Bredas, Phys. Rev. Lett. {\bf 89}, 275503 (2002).
\bibitem{Sak72} J. Sak, Phys. Rev. D {\bf 6}, 3981 (1972).
\bibitem{LLP53} T. D. Lee, F. E. Low, and D. Pines, Phys. Rev. {\bf 90}, 297 (1953).

\end{thebibliography}
\end{document}